\documentclass[aps,pre,twocolumn]{revtex4-2}

\usepackage{graphicx}
\usepackage{dcolumn}
\usepackage{bm}
\usepackage{xcolor} 

\usepackage{longtable}

\begin{document}


\title{A Frenet frame analysis of protein geometry: \\
hints for secondary structure assignments}

\author{M. Prados}
\email{manuprad2003@gmail.com}

\author{M.D. Hernández}%
\email{mdhernandezdelatorre@gmail.com}

\author{F. de Soto}%
\email{fcsotbor@upo.es}
\affiliation{
Depto. Sistemas Físicos, Químicos y Naturales\\
Pablo de Olavide University
}%



\begin{abstract}
This paper deepens into the analysis of the protein secondary structure using Frenet frame to describe the curvature and torsion of the discrete curve formed by the protein $\alpha$-carbons. We show how a simple criterion based on the evaluation of the curvature and torsion of the discrete curve can be useful to pinpoint the presence of some secondary and supersecondary structures in proteins. Moreover, the description of proteins as fixed points of an effective action inspired by an $U(1)$ gauge model is strongly supported by the curvature and torsion observed over a large dataset of proteins in the Protein Data Bank. 

\end{abstract}

\maketitle

\section{Introduction}

Proteins are one of the four major macromolecules that direct life. The protein structure is described at four different levels; the arrangement of amino acids in a polypeptide chain is called its primary structure. The protein secondary structure refers to the way the primary structure of a protein arranges itself as a result of regular
hydrogen bonds forming between the backbone carbonyl and amino groups of each peptide bond. The secondary structure corresponds to local conformations stabilized mainly by hydrogen bonds, such as $\alpha$-helices or $\beta$-strands. The tertiary structure represents the overall three-dimensional folding of a single polypeptide chain, including interactions between secondary structural elements. Finally, the quaternary structure involves the assembly of multiple polypeptide chains into a functional protein complex \cite{creighton1993proteins}.

Knowing the protein structure is crucial for understanding its functions.  Experimental characterization of protein structure is based on techniques such as X-ray crystallography, nuclear magnetic resonance (NMR) spectroscopy, and cryo-electron microscopy (Cryo-EM). These methods provide high-resolution insights into the three-dimensional conformation of proteins \cite{blundell1972protein}.

However, predicting protein's secondary and tertiary structure from its primary structure remains a major scientific challenge due to the immense conformational space and the complexity of intramolecular interactions. Despite remarkable progress, accurately modeling secondary, tertiary and quaternary structures continues to be difficult because of the the dynamic nature of proteins and the influence of the environment \cite{dill2012protein}. 
In recent years, advances in computational biology have transformed the field. Deep learning algorithms such as AlphaFold have dramatically improved the accuracy of structure prediction by leveraging large datasets and sophisticated neural network architectures to predict protein conformations with unprecedented precision \cite{jumper2021highly}.

Secondary structures like $\alpha$-helices and $\beta$-strands are fundamental motifs. In our current understanding of proteins, the assignment of secondary structures relies on the atomic coordinates determined experimentally, using algorithms such as the Dictionary of Secondary Structure in Proteins (DSSP)~\cite{kabsch1983dictionary}, that has become a standard in the field. DSSP identifies structural elements based on hydrogen bond patterns and backbone geometry as characterized by the electrostatic energy between  $C=O$ and $N–H$ groups. Specific bonding patterns define $\alpha$-helices (\textit{i} $\rightarrow$ \textit{i}+4), 3$_{10}$-helices (\textit{i} $\rightarrow$ \textit{i}+3), $\pi$-helices (\textit{i} $\rightarrow$ \textit{i}+5), and $\beta$-sheets (inter-strand bonds). Ramachandran angles ($\phi$, $\psi$) are used as complementary criteria, and a minimum number of consecutive residues must meet the same pattern to be considered a stable secondary structure. 

A large dataset of protein data is stored in the Protein Data Bank (PDB)~\cite{PDB}, including atomic positions for each amino acid in the protein, as well as annotations including the information on the secondary structure of the protein, typically generated using standard algorithms, mainly DSSP. Other methods for assigning secondary structures use geometric data of the protein backbone; for example, King and Johnson \cite{king1999assigning} introduced a geometric approach to secondary structure assignment based on backbone dihedral angles and interatomic distances, rather than hydrogen-bond energies. Their method analyzes characteristic angular and distance patterns along the peptide backbone to classify regions as $\alpha$-helices, 3$_{10}$-helices, $\beta$-strands, or turns.

A different perspective is provided by the Frenet-frame analysis of polymers~\cite{rappaport2007differential,rappaport2006worm,kats2002frenet,giomi2010statistical,castro2019stochastic}. This idea has been applied to proteins by A.J. Niemi and collaborators in a series of papers~\cite{Niemi:2002mv,Danielsson:2009qm,Chernodub:2010xz,Hu:2011wg,Krokhotin2012,Molochkov:2017jmv,Melnikov:2019len,molkenthin2011discrete} where they
have analyzed protein characteristics based on the geometry of the alpha-carbon backbone using the Frenet frame. Following these works, proteins backbones can be understood as links whose degrees of freedom are curvature $\title{\kappa}$ and torsion $\title{\tau}$. This offers a simplified frame to analyze from a purely geometric point of view the polymer structure, where specific secondary structures can be linked 
to the fixed points of a gauge model of proteins, where the curvature inspired in a Higgs-like action.

Our goal in this paper is to use the Frenet frame description of protein backbone geometry to delve into protein structure characterization, showing that this framework can be useful for characterizing secondary structures such as $\alpha$-helices, $\beta$-strands or turns. The structure of the paper is as follows: in Sec.~\ref{sec:frenet} we review three-dimensional curve description in terms of Frenet frame, in Sec.~\ref{sec:results} we apply this frame to analyze typical protein secondary structures, delve into the understanding of this curve as a gauge model in terms of the curvature and torsion of the curve, and propose a method based solely on the geometry of the $\alpha$ carbon backbone structure to identify secondary structures. Finally, we present our conclusions.

\section{Frenet frame\label{sec:frenet}}
\subsection{Frenet frame in the continuum}

Frenet trihedron provides a framework to describe three-dimensional curves in terms of its curvature and torsion, parameters that quantify the variations of the trihedron vectors along the curve. The trihedron is formed by the tangent $\vec{T}$, normal $\vec{N}$ and binormal $\vec{B}$ vectors, whose derivatives along the curve are given by:
\begin{eqnarray}
    \frac{d\vec{T}}{ds} &=& \kappa \vec{N} \label{eq:frenetT}\\ 
    \frac{d\vec{N}}{ds} &=&  -\kappa \vec{T} + \tau \vec{B}\label{eq:frenetN}\\ 
    \frac{d\vec{B}}{ds} &=&  -\tau \vec{N} \label{eq:frenetB}
\end{eqnarray}
that can be compactly written as:
\begin{equation}\label{eq:frenet}
    \frac{d}{ds}\left(\begin{array}{c}
    \vec{T} \\ \vec{N} \\ \vec{B}
    \end{array}\right) = \left(\begin{array}{c c c}
    0 & \kappa & 0\\
    -\kappa &0 & \tau \\
    0 & -\tau  &0 
    \end{array}\right)
    \left(\begin{array}{c}
    \vec{T} \\ \vec{N} \\ \vec{B}
    \end{array}\right)\ .
\end{equation}
Known the curvature and torsion along the curve, $\kappa(s)$ and $\tau(s)$, the curve is uniquely defined in three-dimensional space, up to a rigid rotation.

Frenet frame described by Eq.~\ref{eq:frenet} can be understood as a particular choice of a larger class of trihedrons describing the curve, where the vectors belonging to the orthogonal plane of the curve are written as the complex combinations:
\begin{equation}
    \vec{E}^\theta_{\pm} = e^{i\theta}(\vec{N}\pm i\vec{B})\ .
\end{equation}
The angle of rotation $\theta$ redefines the normal and binormal vectors of the Frenet frame, keeping the curve unchanged, hence redefining the curvature and torsion as $k_\theta=e^{i\theta}\kappa$ and $\tau_\theta=\tau+\frac{d\theta}{ds}$. In this sense, the rotation can be understood as a U(1) gauge transformation, that leaves the curve unchanged. The existence of this gauge symmetry led the authors of \cite{Niemi:2002mv,Danielsson:2009qm,Chernodub:2010xz} to propose a Landau-Ginzburg energy functional $E= \int_0^L ds\ \mathcal{E}(\kappa[s],\tau[s])$:
\begin{equation} \label{eq:contE}
    \mathcal{E}(\kappa,\tau) = \left(\left(\frac{d }{ds}-i\tau_\theta\right)\kappa_\theta\right)^2 + c \left(\kappa_\theta^2-\mu^2\right)^2 + d\tau_\theta+\cdots
\end{equation}
analogous to an Abelian Higgs model where $\kappa_\theta$ and $\tau_\theta$ play the role of scalar and gauge fields respectively.

The curve described by the polymer backbone, such as a protein, is nonetheless discrete, with $C_\alpha$–$C_\alpha$ distances showing a sharp distribution around $3.8$ \AA. We will therefore resort to the description of this discrete curve within the Frenet frame. 

\subsection{Frenet frame for a discrete curve}

The discrete set of positions occupied by the protein's alpha carbons, $C_\alpha$, defines a discrete curve in three-dimensional space that can be described using the Frenet frame. At this level, the protein backbone is simplified to a set of discrete links connecting each pair of consecutive carbons $C_\alpha$. Each link is defined by the angles of curvature and torsion that measure how this link bends with respect to the previous one. The protein can therefore be described as a gauged model of links whose relative orientations give rise to the three-dimensional curve.

The curvature and torsion values of the curve can be straightforwardly computed if the derivatives in Eqs. (\ref{eq:frenetT}-\ref{eq:frenetB}) are approximated by a forward difference as:
\begin{eqnarray}
    \vec{T}_{i+1} &\propto& \vec{T}_i + s_i \kappa_i \vec{N_i} \label{eq:discreteT}\\
    \vec{N}_{i+1} &\propto& \vec{N}_i - s_i \kappa_i \vec{T}_i + s_i \tau_i \vec{B}_i \label{eq:discreteN}\\
    \vec{B}_{i+1} &\propto& \vec{B}_i - s_i \tau_i \vec{N_i} \label{eq:discreteB}\,,
\end{eqnarray}
where $\vec{T}_i$ defines the direction of the vector between the two adjacent positions $i$ and $i+1$ of the discrete curve, $\vec{R}_i$ and $\vec{R}_{i+1}$, i.e.
\begin{equation}
    \vec{T}_i = \frac{1}{s_i}\left(\vec{R}_{i+1}-\vec{R}_i\right)
\end{equation}
and $s_i=|\vec{R}_{i+1}-\vec{R}_i|$ the distance between two consecutive positions. Normal and binormal vectors can then be obtained from the tangent vectors, $\vec{T}_i$, as:
\begin{equation}\label{eq:defB}
    \vec{B}_i = \frac{\vec{T}_{i-1} \times \vec{T}_{i}}{|\vec{T}_{i-1} \times \vec{T}_{i}|}
\end{equation}
and
\begin{equation}\label{eq:defN}
    \vec{N}_i = \vec{B}_{i} \times \vec{T}_{i}\,.
\end{equation}

In Eqs.~(\ref{eq:discreteT}-\ref{eq:discreteB}) the proportionality sign has been used to emphasize that the vectors at position $i+1$ are normalized, and thus in the discrete case ($s\ne0$) a normalization factor is required.

Once the Frenet frame vectors have been computed, one can obtain the dimensionless curvature and torsion that we will define as:
\begin{eqnarray}
    \overline{\kappa}_i &=& s_i\kappa_i = \left(\vec{T}_{i+1} - \vec{T}_i\right) \cdot \vec{N_i} \label{eq:newk}\\
    \overline{\tau}_i &=& s_i\tau_i= \left(\vec{N}_{i+1} - \vec{N}_i\right) \cdot \vec{B_i}\,. \label{eq:newt}
\end{eqnarray}
As $s_i\approx s=3.8$ \AA\ for the vast majority of $C_\alpha-C_\alpha$ links, we will work with the dimensionless curvature and torsion $\overline\kappa$ and $\overline\tau$, and the dimensions of $\kappa$ and $\tau$ are provided by this distance, that sets the typical length of the curve's characteristics. 

It is interesting, before moving on, to consider the meaning of these parameters for a polymer structure. In the case of two consecutive tangent vectors that are almost in the same direction, $\vec{T}_{i+1} \approx \vec{T}_i$, the normal vector at position $i$, $\vec{N}_i$ will be almost orthogonal to the tangent vector at the next one, $\vec{T}_{i+1}$, and thus $\overline\kappa\approx 0$ according to Eq.~(\ref{eq:newk}). The same will occur for tangent vectors that are opposite to each other. That means that small curvatures (in absolute value) will correspond to either non-curved geometries or to very large bending, that will be forbidden by molecules self-avoidance.
Concerning torsion, if four consecutive $C_\alpha$ were contained in the same plane, the two binormal vectors corresponding to two central positions would be identical (orthogonal to that plane) and therefore the torsion would be equal to zero

Note that the dimensionless parameters $\overline\kappa_i$ and $\overline\tau_i$ are closely related to the bond $\psi_i$ and torsion $\theta_i$ angles defined in \cite{Krokhotin2012} (see Eqs.~(21-22) therein):
\begin{eqnarray}
    \psi_i &=& \arccos \left(\vec{T}_{i+1} \cdot \vec{T}_i \right) \label{eq:oldk}\\
    \theta_i &=& \arccos \left(\vec{B}_{i+1} \cdot \vec{B}_i\right) \,. \label{eq:oldt}
\end{eqnarray}

Our definition in Eqs.~(\ref{eq:newk},\ref{eq:newt}) is related to the angles in Eqs.~(\ref{eq:oldk},\ref{eq:oldt})  by $\overline\kappa_i \approx \sin\psi_i$, $\overline\tau_i \approx \sin\theta_i$.
These equations imply that for a curve with $\overline\kappa=0$ ($\overline\tau=0$), also $\psi=0$ ($\theta=0$), while for the case of typical secondary structures, the values of curvature and torsion and their respective bond and torsion angles have been included in Tab.~\ref{tab:values}.

\begin{table}[h]
\centering
\begin{tabular}{|c|c|c|}
\hline
    Structure & Curvature & Torsion \\
\hline
    $\alpha$-helix & $\overline\kappa=\pm 1$ & $\overline\tau=\sin(1)\approx 0.84$ \\
    & ($\psi=\pm\pi/2$) & ($\theta=1$)\\
\hline
$\beta$-strand & $\overline\kappa=\pm\sin(1)\approx \pm 0.84$ & $\overline\tau=0$ \\
& ($\psi=1$)& ($\theta=\pi$)\\
\hline
\end{tabular}
\caption{Values of the curvature $\overline\kappa$ and torsion $\overline\tau$ and the corresponding bond $\psi$ and torsion $\theta$ angles in typical secondary structures in proteins~\cite{Krokhotin2012}.}
\label{tab:values}
\end{table}

The most important differences between the definitions of the bond and torsion angles in Eqs.~(\ref{eq:oldk},\ref{eq:oldt}) and the curvature/torsion of Eqs.~(\ref{eq:newk},\ref{eq:newt}) is that the definition in Eq.~\ref{eq:oldk} is positive definite, while the new definition in Eq.~\ref{eq:newk} isn't. 
As we will see below, the information on the sign can be a rich source of information for identifying secondary structures.

One of the more subtle issues when computing 
curvature and torsion for a discrete set of points are the cases where there is an inflection point, where either the normal or binormal vectors change sense (see Fig. 6 of ref. \cite{Hu:2011wg}). Inflection points are inherently difficult to identify, and lead to ambiguities in the sign of the curvature \cite{Danielsson:2009qm}. Indeed, since it is not well defined, there is no way of deciding whether the result of Eq.~(\ref{eq:defB}) is the most natural choice or a binormal vector in the opposite direction. Both definitions describe the same curve once the curvature and torsion have been computed accordingly using Eqs.~(\ref{eq:newk},\ref{eq:newt}), and the different prescriptions amount to decide between a binormal vector that forms an angle larger or smaller than $\pi/2$ with the preceding one. 
Here, we will adopt the prescription of modifying the definition of $\vec{B}_i$ in (\ref{eq:defB}) altering its sense so that two consecutive vectors always form an angle smaller than $\pi/2$, and subsequently define the normal vector using Eq.(\ref{eq:defN}). This redefinition amounts to a change of sign in the curvature as we reverse the vectors $\vec{N}_i$ and $\vec{B}_i$. This change in the curvature with respect to the direct application of Eqs.~(\ref{eq:newk},\ref{eq:newt}) will last until the next change in the definition of a binormal vector, while for the torsion it amounts to a change of sign for amino acid $i+1$.

Although the sign of the curvature is {\it per se} not well defined, we have found this definition useful for characterizing secondary structures in proteins. For an helix, for example, the curvature remains either positive or negative along the helix except for the rare cases where the helix has a large bending at some point and appears therefore distorted. For $\beta$-strands, on the contrary, the curvature has alternating signs at adjacent positions of the curve, being this characteristic a strong evidence of the presence of $\beta$-strands. As in the case of helices, only in the case of $\beta$-strands with a large bending our choice of the binormal vector direction does not ensure the appearance of this alternating sign of the curvature.

In this work we will proceed as follows: we use the Protein Data Bank to download the pdb file that contains the positions $\vec{R}_i$ of all the atoms in each one of the polypeptide chains that conforms the protein, and compute the Frenet vectors for each position of the chain, flipping the sense of the binormal vector following the criterion explained above. Then we apply our definition for the calculation of the dimensionless curvature and torsion, that will characterize the discrete three-dimensional curve.

\subsection{A gauge model of protein structures}

The gauge model of polymers inspired by the functional form of Eq.~(\ref{eq:contE}) is relatively general, and supports the description of the three-dimensional curves described by protein chains based on their geometric structure, without any reference to the underlying atomic nature of the molecule.

For the discrete chain formed by the $\alpha$-carbons in polypeptide structures, one can resort to a discrete functional energy~\cite{Chernodub:2010xz,Danielsson:2009qm,Hu:2011wg,Krokhotin2012}:
\begin{equation}
E = \sum_i -2 \overline{\kappa}_{i+1}\overline\kappa_i + 2\overline\kappa_i^2 + U(\overline\kappa_i,\overline\tau_i) \label{eq:discE}
\end{equation}    
with $\overline\kappa_i,\overline\tau_i\in [-1,1]$ and where $U(\overline\kappa_i,\overline\tau_i)$ is a generalized potential. This model incorporates first neighbors coupling for $\overline\kappa_i$ and only on-site interactions for $\overline\tau_i$.

For illustrative purposes, we can think of a two-dimensional discrete curve, described only in terms of its curvature, with a Higgs-inspired potential
\begin{equation}
    U^{2D}(\overline\kappa_i) = \frac{1}{4}\lambda \overline\kappa_i^4-\frac{1}{2}\mu^2\overline\kappa_i^2\,.
\end{equation}
It can be easily shown that this discrete model supports i) constant curvature solutions $k_i=k_{i+1}=\pm\sqrt{\frac{\mu^2}{\lambda}}$ corresponding for $\mu\ne 0$ to the minima of the Higgs broken phase in the continuum that would represent a circumference (the 2D counterpart of helices) and ii) alternating sign solutions $k_{i+1}=-k_i=\pm\sqrt{\frac{\mu^2-8}{\lambda}}$ for $\mu^2>8$ corresponding to the typical zig-zag structure of $\beta$-strands.

For the three-dimensional curves, we may devise a functional form of $U(\overline\kappa_i,\overline\tau_i)$ that departs from the two-dimensional case and contains more general terms for the dependence on the torsion, $\overline\tau_i$. We can quite generally write it as an expansion in powers of $\overline\kappa_i$ and $\overline\tau_i$ as: 
\begin{equation}\label{eq:U3D}
U(\overline\kappa_i,\overline\tau_i)=U^{2D}(\overline\kappa_i)+\sum_{n,m=0}^\infty c_{nm} \overline\kappa_i^{2n} \overline\tau_i^{m+1}\,,
\end{equation}
where only even powers in $\overline\kappa_i$ have been included to support the symmetry $\overline\kappa_i\to -\overline\kappa_i$. The lowest order terms of this expansion correspond to $\overline\tau_i$, $\overline\tau_i\overline\kappa^2_i$  (breaking the symmetry in $\overline\tau_i$ and thus supporting solutions with either left-handed or right handed chirality), $\overline\tau^2_i$ (a mass-like term for $\overline\tau_i$),  and $\overline\tau_i^2\overline\kappa^2_i$ (similar to the Higgs mechanism). These terms are presented in Eq.~(3) of \cite{molkenthin2011discrete} as well as in \cite{Danielsson:2009qm, Chernodub:2010xz}, but indeed they are \textit{the} natural extension of the two-dimensional energy functional to three dimensions, as discussed above; they are just the first terms of an expansion in $\overline\kappa_i$ and $\overline\tau_i$. The values of the coefficients $c_{nm}$ of the terms mentioned above, or even higher order terms, can be tuned to support the existence of fixed points corresponding to the secondary structures of proteins or to other possible characteristics of other three-dimensional curves. In particular, for proteins, one can figure out that there are values of the coefficients $c_{nm}$ that support the existence of fixed points of Eq.~(\ref{eq:discE}) for the values of $\overline\kappa_i$ and $\overline\tau_i$ corresponding to the typical secondary structures of proteins such as the ones in Table \ref{tab:values}.

\section{Results\label{sec:results}}

\subsection{Secondary and super-secondary structures}

The application of our prescription for calculating curvature and torsion allows the identification of some of the secondary structures in proteins, as well as the most common transitions among them. The geometric pattern of helices corresponds to $\overline{\kappa}\approx \pm 1$, with a torsion $\overline\tau\approx \sin(1)\approx 0.84$ for $\alpha$-helices and a larger value for $3/10$-helices. For $\beta$-strands, they will be characterized by a small torsion (theoretically zero for a completely flat $\beta$-strand) and a curvature that has alternating signs with $|\overline\kappa|\approx \sin(1)$, according to the values in Table~\ref{tab:values}. 
In this sense, our prescription (\ref{eq:newk}) in conjunction with the definition of the binormal vector that tries to disentangle inflection points will provide  valuable information for identifying the presence of $\beta$-strands, whose typical zigzag structure is associated with alternating signs in the curvature.

\begin{figure}[h]
  \centering
  \begin{tabular}{c c}
     \includegraphics[width=0.572\columnwidth]{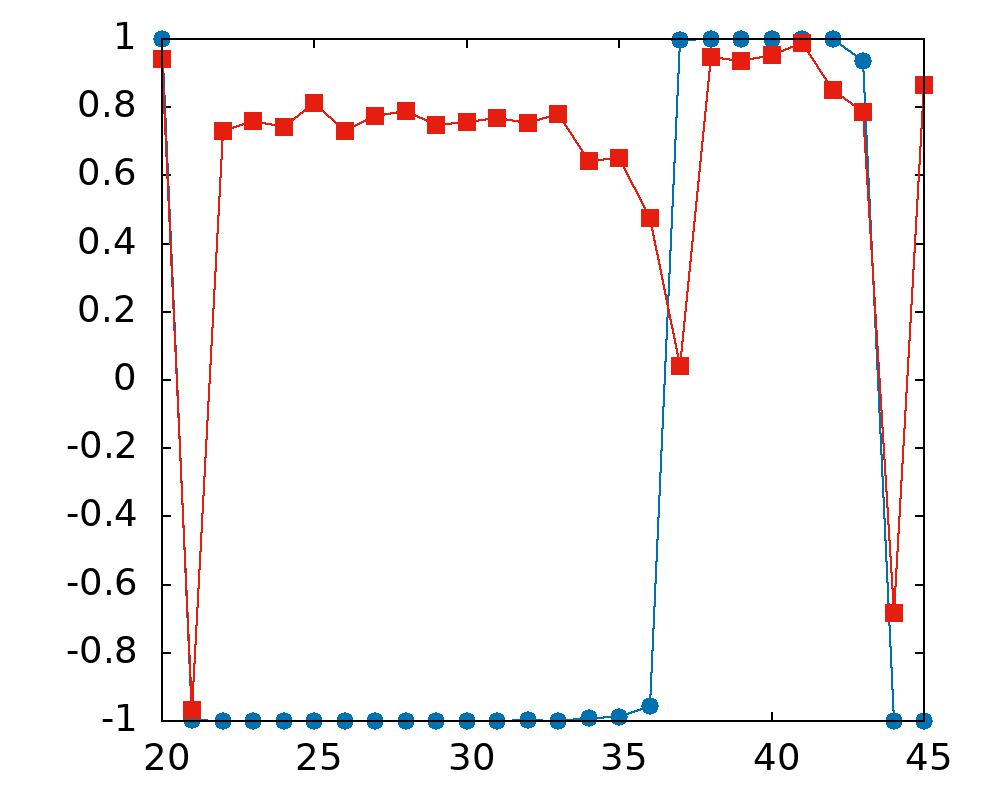}  &  
     \includegraphics[width=0.428\columnwidth]{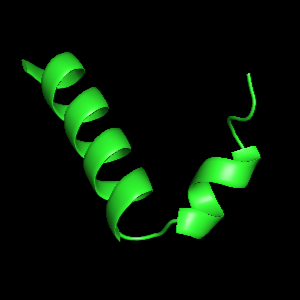} 
  \end{tabular}
  \caption{Transition between an $\alpha$-helix and a $3/10$ helix in protein 1A3N (Deoxy Human Hemoglobin) around amino acid number 36 as characterized in terms of curvature (blue circles) and torsion (red squares) [left] and its 3D counterpart [right].}
  \label{fig:1a3n_structure_and_curvature_torsion}
\end{figure}

\begin{figure}[h]
  \centering
  \begin{tabular}{c c}
     \includegraphics[width=0.572\columnwidth]{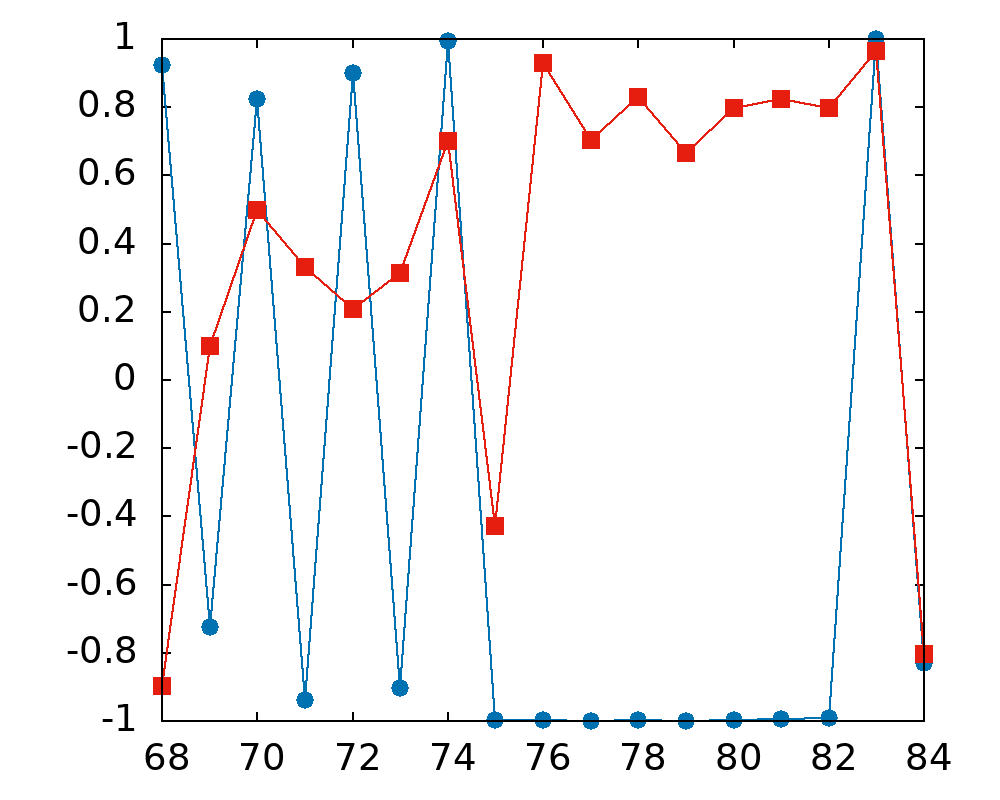}  &  
     \includegraphics[width=0.428\columnwidth]{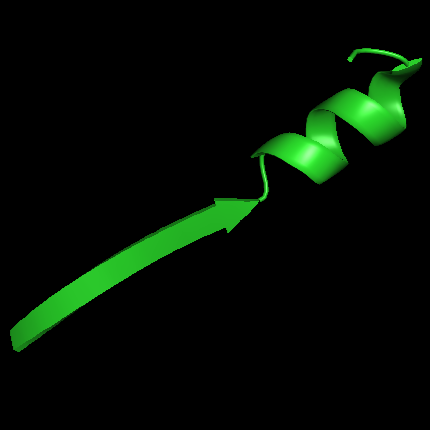} 
  \end{tabular}
  \caption{Transition between a $\beta$-strand and an $\alpha$-helix in protein 2PAB (prealbumin) around amino acid number 75 as characterized in terms of curvature (blue circles) and torsion (red squares) [left] and its 3D counterpart [right].}
  \label{fig:2pab_structure_and_curvature_torsion}
\end{figure}

\begin{figure}[h]
  \centering
  \includegraphics[width=\columnwidth]{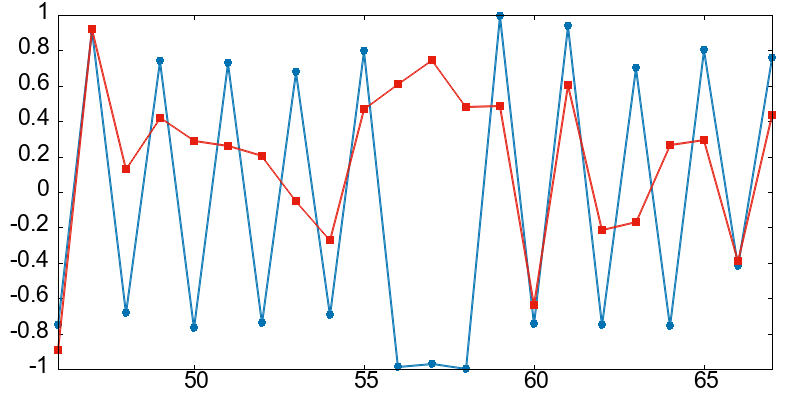}\\
  \includegraphics[width=0.95\columnwidth]{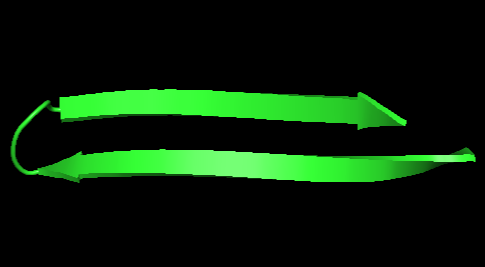}
  \caption{Hairpin transition between two antiparallel $\beta$-strands in protein 3CNA (concanavalin) as characterized in terms of its curvature and torsion (top) and the corresponding three-dimensional structure (bottom).}
  \label{fig:3cna}
\end{figure}

In Figs.~\ref{fig:1a3n_structure_and_curvature_torsion}-\ref{fig:3cna} we show some of the typical secondary and supersecondary structures that appear in proteins, showing both our results for curvature and torsion and the three-dimensional representation of these structures, and they exemplify the main secondary structures found in proteins and some of the transitions among them. 
In Fig.~\ref{fig:1a3n_structure_and_curvature_torsion}, 
we represent the curvature and torsion of the protein Deoxy Human Hemoglobin (with PDB id 1A3N) between residues 20 and 45, according to the PDB labeling. The presence of an $\alpha$-helix spanning residues 21–36 can be clearly identified, characterized by a curvature $\overline{\kappa} \approx -1$ and a positive torsion of about $\overline{\tau} \sim 0.7$. Another helix with $\overline{\kappa}\approx +1$ appears between amino acids 38 and 43. This second helix is characterized by a larger torsion, signaling indeed the fact that this second structure is a $3/10$ helix, as identified in PDB. It is worth noting that the transition between the two helices occurs at residues 36-37 and is remarkably fast, in the sense that one helix starts immediately after the previous one ends, without a smooth transition from one to the other.

Fig.~\ref{fig:2pab_structure_and_curvature_torsion} provides an archetypal example of the transition between a $\beta$-strand and an $\alpha$-helix in prealbumin (protein id 2PAB in PDB). Before residue number 75, the alternating signs of the curvature indicate the presence of a $\beta$-strand, while after residue 75 the curvature becomes constant, with $\overline\kappa=-1$, characteristic of a helix.

Finally, in Fig.~\ref{fig:3cna} we have represented an example of the hairpin transition between two antiparallel $\beta$-strands in the protein concanavalin (3CNA). The two adjacent antiparallel $\beta$-strands are separated by a short turn that connects both strands running in opposite directions, causing the polypeptide chain to bend by approximately \(180^{\circ }\). This short loop is crucial for the formation of structures like $\beta$-hairpins and enables the antiparallel arrangement of the strands. Many such supersecondary structures can be easily found in a search over PDB files. If one observes the curvature of the residues in the transition region, it is quite similar in all of them, and one can find that they adopt a conformation that is compatible with an $\alpha$-helix ($\overline\kappa\approx -1$ and $\overline\tau$ vaguely compatible with $\sin(1)\approx 0.84$).

In all these structures, both the presence of secondary structures and the transitions among them can be characterized in terms of curvature and torsion. The authors of \cite{Niemi:2002mv,Chernodub:2010xz,Hu:2011wg,Krokhotin2012,Molochkov:2017jmv,Melnikov:2019len} studied these transitions as soliton solutions of an energy function as (\ref{eq:discE}) that includes nearest-neighbor interactions for $\overline\kappa$ and on-site interactions for $\overline\tau$. The transitions exemplified by Figs.~\ref{fig:1a3n_structure_and_curvature_torsion}, \ref{fig:2pab_structure_and_curvature_torsion}, and \ref{fig:3cna}, however, seem too abrupt to fit the dependence of the curvature and torsion on the amino acid number. Indeed, in Fig.~\ref{fig:3cna} it seems that the transition region between two $\beta$-strands corresponds to another minimum of the energy functional (\ref{eq:discE}).
The hairpin region in Fig.~\ref{fig:3cna} is characterized by a curvature close to $|\overline\kappa|=1$, showing that even a small loop can adopt this stable configuration even without hydrogen bonds that are responsible for the stability of helices in proteins.

Of particular interest are supersecondary structures such as $\alpha$-helices turns or $\alpha-\alpha$ or $\alpha-3/10$ transitions such as the one shown in Fig.~\ref{fig:1a3n_structure_and_curvature_torsion}, $\alpha-\beta$ transitions such as the one represented in Fig.~\ref{fig:2pab_structure_and_curvature_torsion} or the $\beta$-hairpin structure between antiparallel $\beta$-strands shown in Fig.~\ref{fig:3cna}. It is important to note that identifying these structures in terms of curvature and torsion is made possible by our definition in Eqs.~(\ref{eq:newk}-\ref{eq:newt}), that provides additional information with respect to Eqs.~(\ref{eq:oldk}-\ref{eq:oldt}) associated with the sign of the curvature. 
In the present prescription, in any case, the curvature–sign ambiguity still persists, but only in the cases of $\beta$-strands that suffer a large bending at some point or bended helices the results deviate from the ones presented above.

\subsection{Distribution of curvature and torsion}

\begin{figure}[h]
  \centering
  \includegraphics[width=\columnwidth]{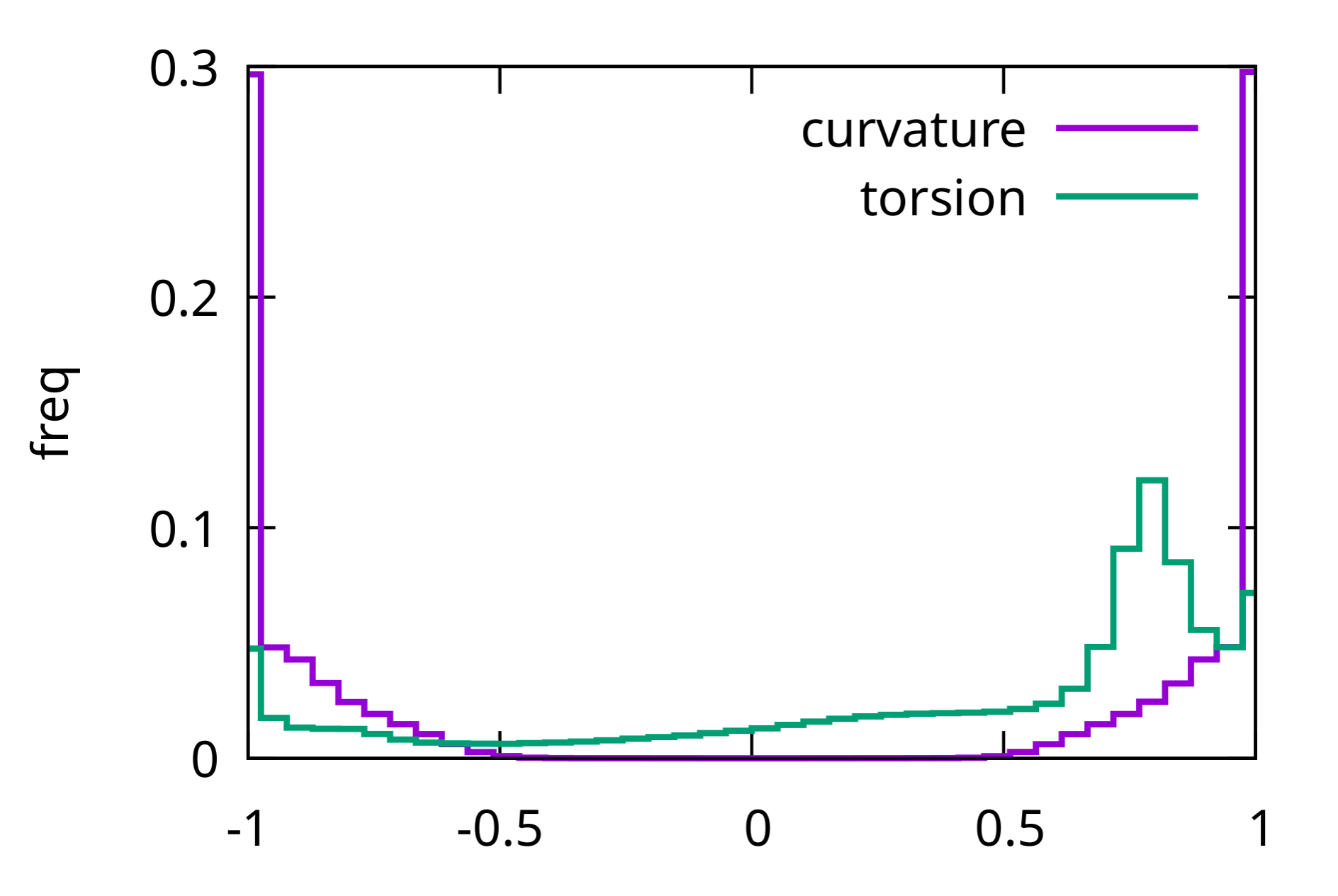}
  \caption{Histogram of the values of curvatures and torsion obtained over a large number of randomly chosen proteins from PDB.} 
  \label{fig:histograma_curvaturas}
\end{figure}

The $(\overline\kappa,\overline\tau)=(\pm 1,\sin 1)$ configuration seems to be the most stable combination in proteins. This is evidenced by the following analysis; we have randomly downloaded over 18000 protein structures from PDB, and computed the curvature and torsion for each of the polypeptide chains contained in these files. The histogram of the curvature and torsion values obtained has been represented in Fig.~\ref{fig:histograma_curvaturas}. These histograms show several interesting characteristics: i) the distribution of curvatures is symmetric in $\overline\kappa$, as expected, and ii) almost $60\%$ of the values correspond to $\overline\kappa=\pm 1$ which is typical of helices. The large fraction of values with a curvature close to $|\overline\kappa|=1$ manifests that this configuration is very stable and should correspond, in Eq.~(\ref{eq:discE}), to its lowest energy minimum. 

The histogram of torsion values in Fig.~\ref{fig:histograma_curvaturas} is asymmetric, in contrast to the curvature distribution, showing a preference for positive values of the torsion. A distinct peak at positive values of $\overline\tau$ shows that about $40\%$ of the values are contained between $\overline\tau=0.65$ and $\overline\tau=0.95$, compatible with the values of $\alpha$-helices according to the values in Table \ref{tab:values}. In any case, the fact that the probability of finding curvature $|\overline\kappa|\approx 1$ is much larger than $\overline\tau\approx \sin 1$ would indicate that the curve tends to have a large curvature even in geometries that are not helices.

The signature of $\beta$-strands in Fig.~\ref{fig:histograma_curvaturas} is less clear than for helices. They would appear as the tails for $\overline\kappa$ slightly larger than $-1$ or slightly smaller than $+1$. In terms of torsion, they would correspond to the region of small $\overline\tau$, although this region does not exhibit a distinct peak but rather a smooth distribution.

One more characteristic emerges from Fig.~\ref{fig:histograma_curvaturas}, and is the existence of peaks at $\overline\tau=-1$ and $\overline\tau=+1$. These peaks are associated with kinks in the curve, where there is a strong bending of the protein, and appear mainly in the turn among different secondary structures.

\subsection{A criterion for identifying secondary structures}

Based on the observation that helices are typically characterized by a curvature $\overline\kappa=\pm 1$ and large torsion, one can devise a criterion to identify the presence of such structures based solely on their geometrical features: curvature and torsion. We will adopt the criterion that a  helix  (either $\alpha$ or $3/10$) is present whenever the amino acid chain is characterized by three or more amino acids with curvature $|\overline\kappa| \in (0.95,1)$ and torsion $\overline\tau>0.5$. Although $3/10$ helices are characterized by larger torsion values than $\alpha$-helices, as illustrated in Fig.~\ref{fig:1a3n_structure_and_curvature_torsion}, it is difficult to establish a criterion to distinguish between them, and therefore we will analyze both types of helices together. For $\beta$-strands, our criterion requires that $\overline\kappa$ alternates in sign over four or more consecutive amino acids, with an average curvature (in absolute value) of $\langle |\overline\kappa|\rangle > 0.6$, where $\langle\cdots\rangle$ stand of the average value. With respect to the torsion, for a flat  $\beta$-strand it would have a vanishing value, and in our criterion we will accept $\beta$-strands when the average value of the torsion satisfies: $|\langle\overline\tau\rangle|<0.5$ in the section of the chain with alternating curvature sign. This criterion for the torsion forbids the identification of strands with a large bending, but in any case these strands are difficult to identify because our criterion for detecting inflection points also fails, and we do not have alternating sign curvature.

These criteria have been established by analyzing a small group of proteins, and later 
applied to the large statistical set of proteins downloaded randomly from Protein Data Bank and used to produce Fig.~\ref{fig:histograma_curvaturas}, consisting of over $70k$ polymer chains from $\sim 18k$ proteins. We have checked whether our criteria based solely on the curvature and torsion of the $C_\alpha$ backbone are capable of detecting the presence of the main secondary structures in proteins. The result is that our criterion identifies correctly over $92\%$ of helices and $\beta$-strands labeled in PDB files, at the time we characterize other ones that are not identified in PDB files. This high rate of positive identification shows that our definition of curvature and torsion in Eqs.~(\ref{eq:newk}-\ref{eq:newt}) can be useful for characterization of polymer three-dimensional structure, enabling a simple and fast recognition of common secondary structures. The identification of less common secondary structures as well as a refinement of our criterion for detecting inflection points, taking into account the identification of secondary structures, is left for a future analysis.

From a different point of view, and in order to assess the quality of the proposed criterion, we present in table \ref{tab:comparison} the percentage of amino acids belonging to helices (either $\alpha$ or $3/10$) and $\beta$-strands for each one of the proteins collected in \cite{king1999assigning}. These percentages are compared with the results reported by the authors and with the corresponding identification in the Protein Data Bank. Our method is at least as good as the one in \cite{king1999assigning}, which considers not only the $C_\alpha$ positions, but also geometric parameters derived from backbone dihedral angles and interatomic distances. 
In their study, they defined secondary structures by analyzing five geometric parameters for each residue: two dihedral angles, $\zeta(i)$, which describes the orientation between consecutive carbonyl groups, and tau $\tau(i)$, which describes the relative position of neighboring $C_\alpha$ atoms; and three interatomic distances, O$(i)$–N$(i+3)$, O$(i)$–N$(i+4)$, and C$(i)$–N$(i+3)$. These values capture the local conformation and potential hydrogen-bonding patterns of the peptide backbone. Based on specific numerical ranges of these parameters, the algorithm classifies residues into $\alpha$-helices, 3$_{10}$-helices, $\beta$-strands and turns. 

Although individually for each protein, both the results of Ref.~\cite{king1999assigning} and our results differ from the PDB assignments, our criterion being simple and with only the information on the $C_\alpha$ positions, seems to be as good as theirs. Overall, our results differ approximately in $8\%$ for helices and $10\%$ for $\beta$-strand localization with respect to the data consigned in PDB, slightly better than theirs. On the other hand, they are able to distinguish between $\alpha$ and $3/10$-helices, while based solely on the torsion of the curve, we have difficulties for setting a quantitative criterion for their distinction.
In both cases, the comparison is made with respect to the helices and strands labeled in PDB files. It is important to remark that some secondary structures may not appear in PDB files.

\begin{widetext}
\begin{center}
\begin{table}[!h]
\centering
    \begin{tabular}{|c|c|c|c|c|c|c|c|}
    \hline
        & & \multicolumn{3}{|c|}{Helices} &\multicolumn{3}{|c|}{Strands}  \\
    \hline
        Protein & ID &  PDB & KJ & This work  & PDB & KJ & This work\\
    \hline
       Azurin   & 5azu  & 24.8 & 17.8 & 8.0  & 31.6 & 33.7 & 38.8 \\
    \hline
        Bence Jones   & 1rei  & 4.8 & 0.0 & 0.0  & 46.2 & 33.5 & 50.2 \\
    \hline
         $\alpha$-Chymotrypsin-a   & 5cha  & 11.3 & 11.0 & 8.8  & 45.1 & 15.8 & 38.1 \\
    \hline
        Concanavalin-a   & 1nls  & 7.3 & 2.6 & 1.3  & 44.0 & 36.4 & 59.4 \\
    \hline
        Cytochrome-c   & 5cyt  & 51.0 & 41.7 & 38.0  & 0.0 & 7.7 & 10.5 \\
    \hline
        Elastase   & 1esa  & 7.6 & 10.1 & 7.6  & 50.6 & 21.3 & 37.1 \\
    \hline
        Flavodoxin   & 5fx2  & 39.6 & 30.8 & 38.2  & 31.9 & 17.1 & 33.0 \\
    \hline
        Hemerythrin   & 1hmd  & 63.6 & 71.2 & 69.1  & 0.0 & 3.6 & 8.4 \\
    \hline
        Hemoglobin   & 2mhb  & 77.6 & 74.6 & 75.8  & 0.0 & 0.0 & 1.8 \\
    \hline
        Lactate dehydrogenase   & 6ldh  & 38.0 & 41.9 & 43.3  & 21.8 & 14.0 & 22.2 \\
    \hline
        $\beta$-Lactoglobulin   & 1beb  & 17.3 & 12.5 & 10.1  & 37.6 & 33.6 & 53.1 \\
    \hline
        Lysozyme   & 4lzt  & 43.7 & 41.4 & 38.1  & 4.8 & 3.9 & 15.1 \\
    \hline
        Myoglobin   & 1vxf  & 80.7 & 80.9 & 76.7  & 0.0 & 0.0 & 2.7 \\
    \hline
        Papain   & 9pap  & 26.3 & 28.9 & 22.9  & 20.5 & 15.2 & 16.6 \\
    \hline
        Pepsinogen   & 3psg  & 31.1 & 17.3 & 10.5 & 14.9 & 27.1 & 41.6 \\
    \hline
        Prealbumin   & 2pab  & 8.1 & 8.5 & 6.3  & 50.5 & 35.3 & 63.3 \\
    \hline
        Ribonuclease-a   & 1rbx  & 21.5 & 22.0 & 21.5  & 29.8 & 19.1 & 33.1 \\
    \hline
        Superoxide dismutase   & 1sxn  & 6.4 & 4.3 & 3.0  & 36.5 & 24.0 & 39.2 \\
    \hline
        T4 lysozyme   & 5lzm  & 67.3 & 66.8 & 65.4  & 11.3 & 6.1 & 3.8 \\
    \hline
        Thermolysin   & 1lnf  & 41.2 & 39.8 & 38.7  & 16.6 & 13.6 & 24.0 \\
    \hline
        Triose phosphate isomerase   & 1ypi  & 54.5 & 39.5 & 40.4  & 18.9 & 14.8 & 23.2 \\
    \hline
       Trypsin   & 1tng  & 5.1 & 11.9 & 0.0 &  37.8 & 19.3 & 46.9 \\
    \hline
    \end{tabular}
    \caption{Percentage of amino acids in the protein that belong either to helices or $\beta$-strands, as given by PDB labelling, ref.~\cite{king1999assigning} and this work. \label{tab:comparison}}
\end{table}
\end{center}
\end{widetext}

\subsection{Relation to the gauge model}

In a series of papers \cite{Danielsson:2009qm,Chernodub:2010xz,Hu:2011wg}, the secondary structures and transitions between them have been studied as stationary points of the energy functional in Eq.~(\ref{eq:discE}). Indeed, for any general form of $U$, the application of Euler-Lagrange equations to this discrete case leads to:
\begin{equation}
    \overline\kappa_{i+1}+\overline\kappa_{i-1}-2\overline\kappa_i = \frac{1}{2}\frac{dU_i}{d\overline\kappa_i} = {\overline\kappa_i}\frac{dU_i}{d\overline\kappa_i^2} \label{eq:DEL}     
\end{equation}
and
\begin{equation}
    \frac{dU_i}{d\overline\tau_i}=0\,, \label{eq:DEL2}
\end{equation}
where, for simplicity, we wrote $U_i\equiv U(\overline\kappa_i,\overline\tau_i)$.

The authors of \cite{molkenthin2011discrete} used the discrete Euler-Lagrange Equations (\ref{eq:DEL},\ref{eq:DEL2}) to write the equivalent of a discrete non-linear Schr\"odinger Hamiltonian in $\overline\kappa$ that is later used to describe transitions among secondary structures as solitons. Our approach here is radically different and somehow complementary; we will instead use Eq.~(\ref{eq:DEL}) to dig into the dependency of $U(\overline\kappa,\overline\tau)$ with curvature and torsion and the determination of the secondary structure of amino acid chains.

\begin{figure}[h]
  \centering
  \includegraphics[width=\columnwidth]{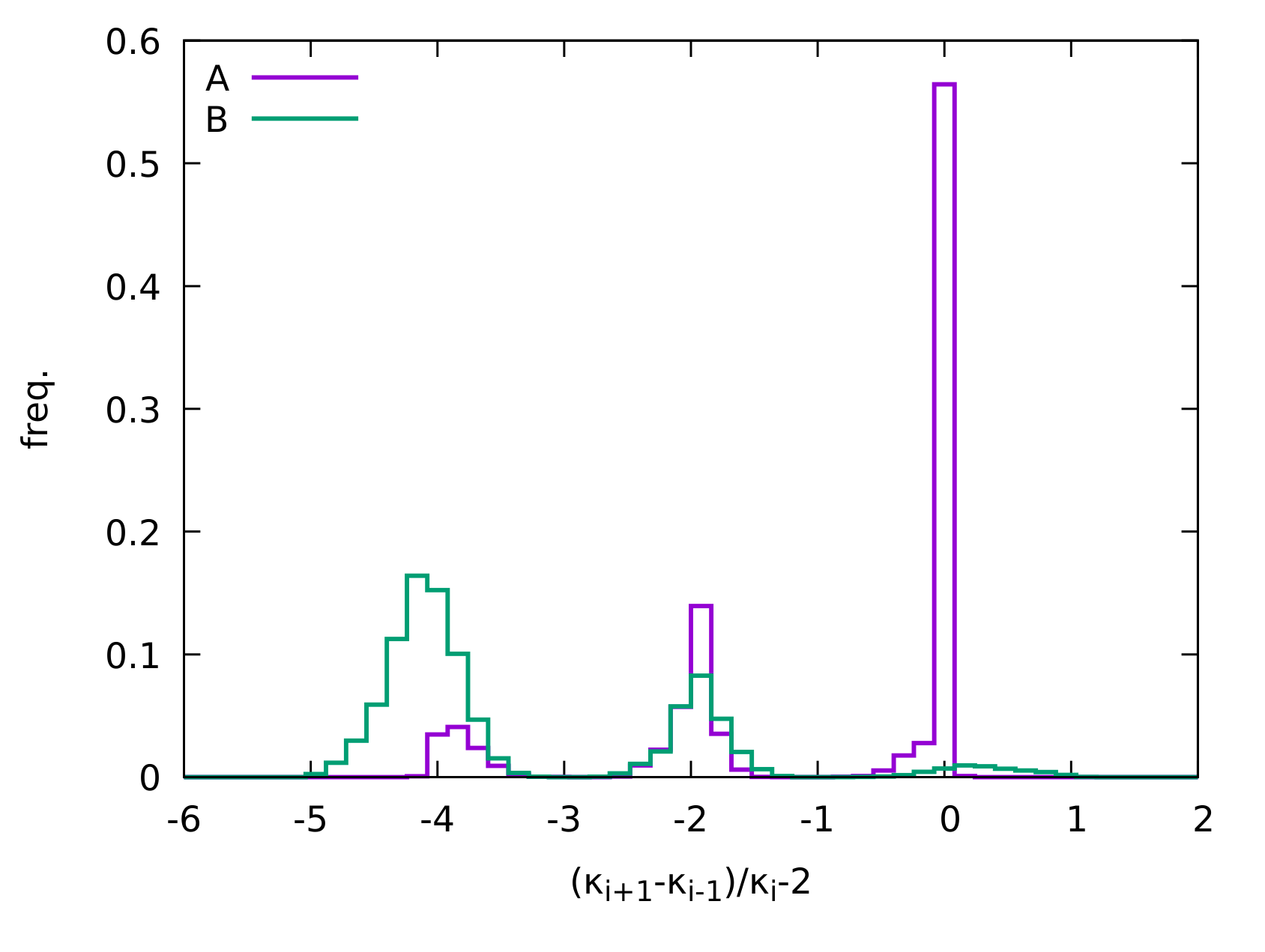}
  \caption{Histogram of the values of $(\overline\kappa_{i+1}+\overline\kappa_{i-1})/\overline\kappa_i-2$ for residues with $|\overline\kappa|>0.95$ (set A) and for  $|\overline\kappa|\in (0.65,0.90)$ (set B).} 
  \label{fig:derivative}
\end{figure}

We will exploit here the set of over 18000 proteins used in the previous section to dig into the dependency of the energy functional with $\overline\kappa$ and $\overline\tau$. To this aim  we have plotted in Fig.~\ref{fig:derivative} a histogram of the values of the quantity:
\begin{equation}\label{eq:dU}
    \frac{\overline\kappa_{i+1}+\overline\kappa_{i-1}}{\overline\kappa_i} - 2 
\end{equation}
that, according to Eq.~(\ref{eq:DEL}), would be equal to $\frac{dU_i}{d\overline\kappa_i^2}$. In this analysis, therefore, we do not \textit{assume} the form of Eq.~(\ref{eq:discE}) but rather will investigate whether this functional form for the energy is reasonable and the shape of $U$
in the hypothesis that the protein backbone curves result as fixed points of Eq.~(\ref{eq:discE}). By selecting the positions of the curve characterized by $|\overline\kappa|>0.95$ (set A), we observe that in most cases $\frac{dU}{d\overline\kappa^2}\approx 0$, indicating that these points correspond to a minimum of the energy functional, such as the one in the Higgs broken phase. 

In terms of curvature, we can consider three possible scenarios for residues with approximate constant $|\overline\kappa|$:
\begin{itemize}
    \item $\overline\kappa_{i-1}=\overline\kappa_{i}=\overline\kappa_{i+1}$, as in the case of an helix, which would imply $\frac{dU_i}{d\overline\kappa_i^2}=0$,
    \item $\overline\kappa_{i-1}=-\overline\kappa_{i}=\overline\kappa_{i+1}$, corresponding to an oscillating curvature, for which $\frac{dU_i}{d\overline\kappa_i^2}=-4$; or
    \item $\overline\kappa_{i-1}=\overline\kappa_{i}=-\overline\kappa_{i+1}$, i.e., where a sign change occurs, such as at the end of an helix, leading to $\frac{dU_i}{d\overline\kappa_i^2}=-2$.
\end{itemize}

Observing the plot corresponding to set A in Fig.~\ref{fig:derivative}, one can appreciate how the data show the appearance of three peaks, as expected, with the following values of $\frac{dU_i}{d\overline\kappa_i^2}$: $0$, which is by large the most frequent value, and indicating that we are at the minimum of the action; $-2$ which may appear, for example, at the end of an helix; and $-4$, where there is an oscillating sign, that may appear in strongly curved $\beta$-strands such as collagen. 

For the data with $|\overline\kappa|\in (0.65,0.90)$ (set B), which would correspond (at least in some cases) to $\beta$-strands, the situation is quite different: they do not correspond to minima of $U$, as the number of points with $(k_{i+1}+k_{i-1})/k_i-2\approx 0$ is rather small, and most of them correspond to the oscillating case described above, for which we have $(k_{i+1}+k_{i-1})/k_i-2\approx -4$. 

The meaning of these results has to be understood as follows: if we assume the three-dimensional curves drawn by proteins as stationary points of the energy functional of the general form of Eq.~(\ref{eq:discE}), there are stationary points corresponding to the minima of the on-site term, $U(\overline\kappa,\overline\tau)$, as appears to be the case for helices, while other stationary points are dynamical, associated with the kinetic term in Eq.~(\ref{eq:discE}), which seems to describe $\beta$-strands. It is important to emphasize that these findings are independent of the functional form chosen for $U(\overline\kappa,\overline\tau)$.

The cases with $\frac{dU_i}{d\overline\kappa_i^2}=-2$ in Fig.~\ref{fig:derivative} would correspond either to a transition between the two minima of the action or to a position where our criterion for the sign correction failed and corresponds to an inner position of an helix or a $\beta$-strand with a large bending. Naturally, we have only analyzed the most frequent structures in proteins, and the patterns that we found might also correspond to other less frequent secondary structures such as $\pi$-helices, polyproline, $\Omega$-loops, etc.

\begin{figure}[h]
  \centering
  \includegraphics[width=\columnwidth]{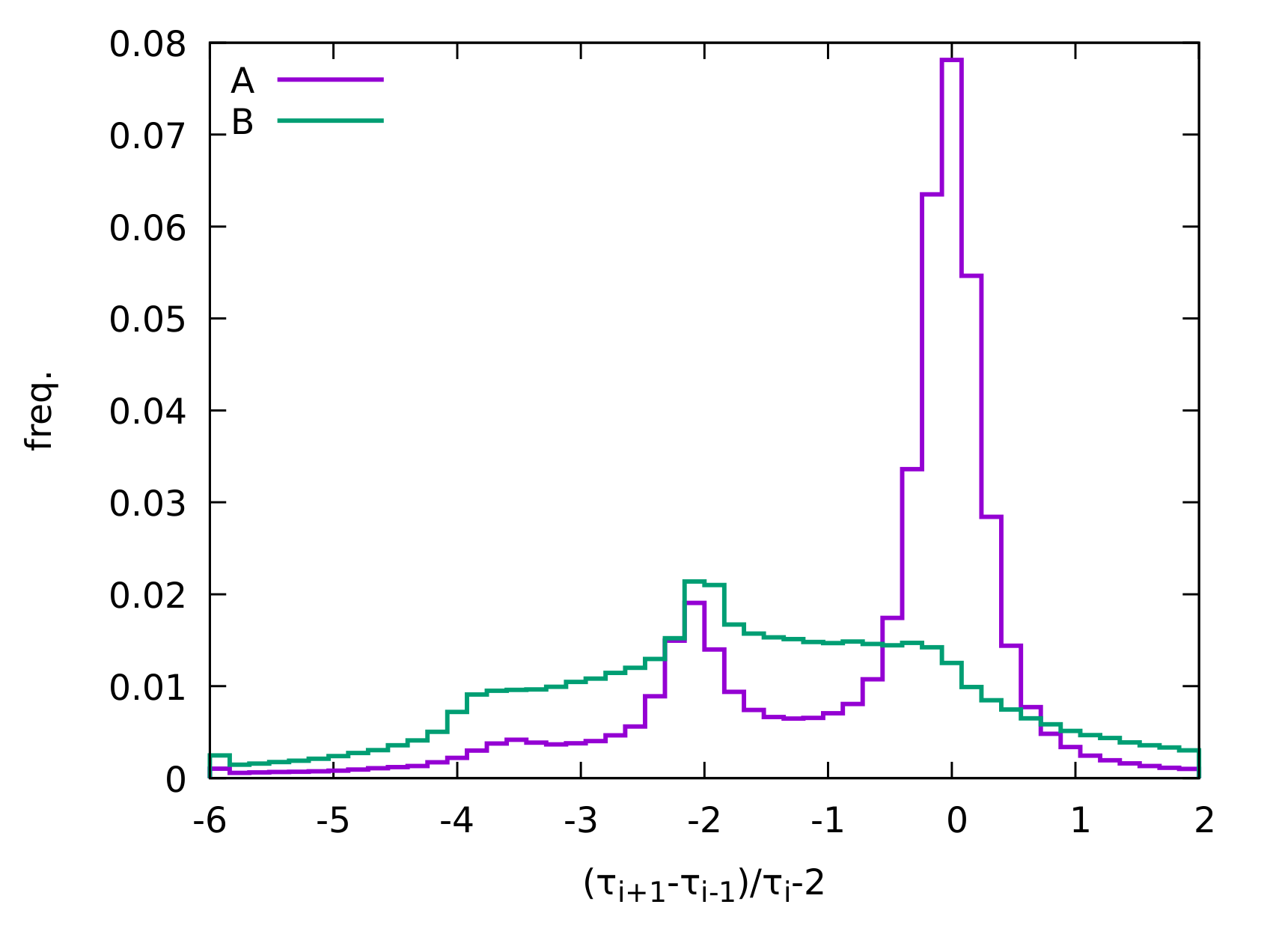}
  \caption{Histogram of the values of $(\overline\tau_{i+1}+\overline\tau_{i-1})/\overline\tau_i-2$ for residues with $|\overline\kappa|>0.95$ (set A) and for  $|\overline\kappa|\in (0.65,0.90)$ (set B).} 
  \label{fig:derivativet}
\end{figure}

Concerning the $\overline\tau$ dependence, an analogous plot has been included in Fig.~\ref{fig:derivativet} for $(\overline\tau_{i+1}+\overline\tau_{i-1})/\overline\tau_i-2$. In this case, for set A, corresponding to a large curvature, we observe a maximum around $(\overline\tau_{i+1}+\overline\tau_{i-1})/\overline\tau_i-2=0$, which corresponds to $\frac{dU_i}{d\overline\tau_i^2}=0$, according to the interpretation in terms of the action (\ref{eq:contE}) or its discrete version of Eq.~(\ref{eq:discE}). Although the peaks are not as clear as in Fig.~\ref{fig:derivative}, some structures can still be observed at $(\overline\tau_{i+1}+\overline\tau_{i-1})/\overline\tau-2\approx -4$ and, more prominently at $(\overline\tau_{i+1}+\overline\tau_{i-1})/\overline\tau-2\approx -2$. Based on the histograms obtained for sets A and B (containing protein helices and $\beta$-strands), this plot appears to exclude the presence of a kinetic term in the action for the torsion $\overline\tau$, as there seems to be no dynamical fixed point of the action characterized by $(\overline\tau_{i+1}+\overline\tau_{i-1})/\overline\tau_i-2\approx -4$, as in the case of the curvature. 

Finally, based on the observed protein backbone structures in nature, a description based on Eq.~(\ref{eq:discE}), with a potential term drawn from scratch as
the one proposed in Eq.~(\ref{eq:U3D}, that we can rewrite as
\begin{equation}
U[\overline\kappa,\overline\tau] = \sum_{m,n} c_{m,n}\overline\kappa^{2m}\overline\tau^n    
\end{equation}
seems to be adequate for describing the polymer structure. The values of the coefficients $c_{m,n}$ are determined by the interactions of the proteins and the structure of the monomers, and knowing them would help drawing the three-dimensional curves adopted by proteins.

\section{Conclusions}

The geometric pattern drawn by $\alpha$-carbons in proteins has been used to compute the curvature and torsion along the curve in a discrete Frenet frame approach. We have proposed an algorithm for computing such values that provides valuable information for identifying and understanding some of the most frequently found secondary structures in proteins and the transitions among them. The information on curvature and torsion has served as a primary indicator of secondary structures, which has proven useful compared to more sophisticated methods involving hydrogen bonds and Ramachandran angles $(\phi,\psi)$. Overall, the simple method proposed is able to recognize most of the helices and $\beta$-strands labeled in the Protein Data Bank files.

We have analyzed the backbone curve drawn by protein's $\alpha$ carbons as the result of a $U(1)$ gauge model, whose energy depends solely on the dimensionless curvature and torsion computed. From the analysis of a large collection of proteins, we have justified the presence of a kinetic term for the curvature, while the kinetic term for the torsion may be absent. Overall, this analysis makes it possible to write an energy functional in terms of curvature and torsion, which includes a kinetic term for the curvature and a potential term and Higgs terms, as well as terms that break the symmetry between positive and negative values of the torsion.

The picture of proteins as fixed points of a functional action in terms of its curvature and torsion, and whose functional form appears in a natural way, is an interesting starting point to analyze the properties of proteins that can be described based on this assumption, and will be the scope of a forthcoming publication.

\end{document}